\documentclass[showpacs,twocolumn,pre]{revtex4}
\usepackage{psfrag,epsfig,amsfonts,amssymb,amsmath,graphicx,slashbox}
\usepackage{dcolumn}

\begin{document}

\title{Weak disorder strongly improves the selective enhancement of 
diffusion in a tilted periodic potential}

\author{Peter Reimann}
\affiliation{Universit\"at Bielefeld, Fakult\"at f\"ur Physik, 33615 Bielefeld, Germany}
\author{Ralf Eichhorn}
\affiliation{Universit\"at Bielefeld, Fakult\"at f\"ur Physik, 33615 Bielefeld, Germany}
\
\begin{abstract}
The diffusion of an overdamped Brownian particle
in a tilted periodic potential is known to exhibit 
a pronounced enhancement over the free thermal diffusion 
within a small interval of tilt-values. 
Here we show that weak disorder in the form of small, time-independent 
deviations from a strictly spatially periodic potential may further boost this 
diffusion peak by orders of magnitude.
Our general theoretical predictions are in excellent 
agreement with experimental observations.
\end{abstract}


\maketitle 

Diffusion plays a key role for mixing and homogenization
but also for particle selection and separation tasks
\cite{gen}.
A particularly simple and common way to
manipulate the force-free diffusion of a 
Brownian particle is by means of a
spatially periodic force field \cite{lif62}
with a non-vanishing systematic component
\cite{fau95,lee06,bli07,evs08,par97,cos99,rei01,diff1,atoms,dna},
i.e. the force derives from
a tilted/biased periodic potential.
Such dynamics arise in a large variety
of different physical systems
\cite{par97,cos99,rei01,diff1}, 
for example colloidal particles in optical 
potentials
\cite{fau95,lee06,bli07,evs08}
cold atoms in optical lattices \cite{atoms},
or globular DNA in microstructures \cite{dna}.
While the force-free thermal diffusion of an 
overdamped Brownian particle is always reduced 
when switching on an unbiased periodic potential
\cite{lif62},
the diffusion coefficient as a function of
an additional bias exhibits a pronounced peak
\cite{par97,cos99,rei01}
in a small vicinity of the so-called critical
tilt, i.e. the threshold bias at which
deterministic running solutions set in.
This theoretical prediction has recently been confirmed 
by several experimental works \cite{lee06,bli07,evs08}.
However, for many experimental realizations of the 
above mentioned large variety of systems involving 
a tilted periodic potential
\cite{fau95,lee06,bli07,evs08,par97,cos99,rei01,diff1,atoms,dna}, 
small, time-independent deviations of the potential
from strict spatial periodicity are practically unavoidable.
The objective of our present Letter 
is a detailed theoretical understanding of 
such weak disorder effects. 
An immediate first guess is that they 
will somehow ``wash out'' the diffusion 
peak around the critical tilt.
For instance, one might argue that 
enhanced diffusion requires a tilt close 
to criticality \cite{rei01} and 
this fine-tuning will unavoidably be 
spoiled by the random variations superimposed 
to the original periodic potential.
Here, we show that exactly the 
opposite is the case: 
Tiny deviations from spatial periodicity
result in an even more pronounced 
peak of the diffusion coefficient.
Hence, the often unavoidable weak disorder 
is not an experimental nuisance but rather
a new tool for sorting particles by way of 
a very strong and selective diffusion 
enhancement for certain species within a 
mixture.
We remark that diffusion in the presence of
temporal rather than spatial disorder 
represents a related but still different case.
It also may result in accelerated diffusion, 
but, in contrast to our present case, already
without a bias \cite{cou07}.

Our starting point is the usual overdamped
Brownian motion in 1D 
\cite{lif62,fau95,lee06,bli07,evs08,par97,cos99,rei01,diff1}:
\begin{equation}
\eta \dot x(t) = -U'(x(t)) + \sqrt{2 \eta kT}\, \xi(t) \ ,
\label{1}
\end{equation}
where $\eta$ is the viscous friction coefficient, 
and thermal fluctuations are modeled by unbiased, 
$\delta$-correlated Gaussian noise $\xi (t)$ 
with thermal energy $kT$.
The potential $U(x)$ consist of a tilted periodic 
part $V(x)$ and ``random'' deviations
$W(x)$ (quenched disorder),
\begin{eqnarray}
& & U(x) = V(x)+W(x)\ ,\ \ V(x)=V_0(x)-xF\ ,
\nonumber
\\
& & V_0(x+L) = V_0(x)\ , 
\label{2}
\end{eqnarray}
where $F$ is a tilting force (static bias) and $L$ 
the spatial period.
Without loss of generality we focus on potentials
``tilted to the right'', i.e. $F \geq 0$.
The quantities of main interest are drift (average velocity)
and diffusion, 
\begin{equation}
v := \lim_{t\to\infty} \frac{\langle x(t)\rangle_\xi}{t} 
\ , \ \ 
D := \lim_{t\to\infty} 
\frac{\langle x^2(t)\rangle_\xi -\langle x(t)\rangle_\xi^2}{2t} \ ,
\label{3}
\end{equation}
where $\langle\cdot \rangle_\xi$ indicates an 
average over the noise $\xi(t)$ in (\ref{1}).
While many of the following considerations can be
generalized to other types of disorder $W(x)$, we
focus on the analytically most convenient case of
{\em unbiased, homogeneous Gaussian disorder}.
In other words, considering $x$ as ``time'',
$W(x)$ is a stationary, Gaussian stochastic
process with mean value $\langle W(x)\rangle=0$ 
and  correlation
\begin{equation}
c(x-y):=\langle W(x)W(y)\rangle \ .
\label{4}
\end{equation}
For simplicity only, we further assume that the
correlation $c(x)$ is monotonically decreasing 
for $x\geq 0$ from
\begin{equation}
\sigma^2 := \langle W (x)^2\rangle = c(0)
\label{5}
\end{equation}
to $c(x\to\infty)=0$. 
A simple example is the critically damped harmonic oscillator
\begin{equation}
\lambda^2\, W''(x) = -2\lambda \, W'(x) - W(x) + 2\sigma\lambda^{1/2}\,\gamma(x)
\label{6}
\end{equation}
driven by $\delta$-correlated Gaussian noise $\gamma(x)$,
yielding a Gaussian $W(x)$ with
$c(x)=\sigma^2(1+|x|/\lambda)\, e^{-|x|/\lambda}$, see Fig. 1.

\begin{figure}
\epsfxsize 1.0 \columnwidth
\epsfbox{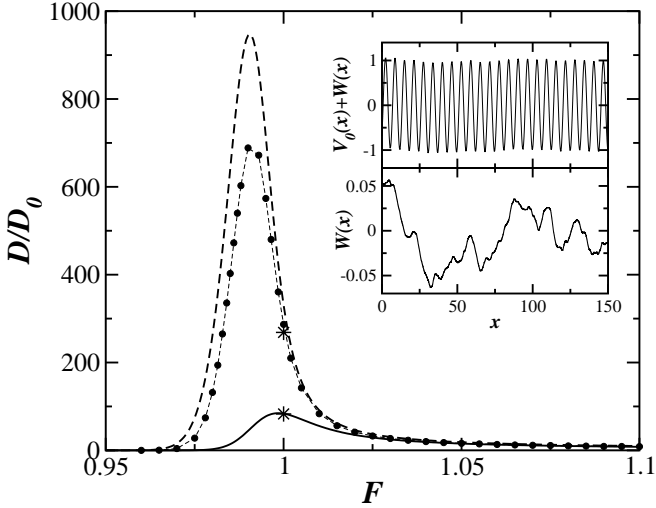}
\caption{
Diffusion $D$ from (\ref{3}) 
versus tilt $F$ for the dynamics 
(\ref{1}), (\ref{2}) with $V_0(x)=\sin(x)$
(i.e. $L=2\pi$, $F_c=1$), $\eta=1$, $kT=D_0=0.001$
(dimensionless units).
The disorder $W(x)$ is a stationary 
Gaussian process satisfying (\ref{6}) 
with $\sigma=0.03$ and $\lambda = L$.
The insets illustrate $U(x,F=0)=V_0(x)+W(x)$ 
(randomness hardly visible) and $W(x)$ (randomness
alone) from (\ref{2}).
Filled dots (connected by short dashes): 
Numerically exact results.
Dashed line: Analytical approximation (\ref{7}),
(\ref{15}), (\ref{17}) with (\ref{22}), (\ref{23}),
and $a=b=0$.
Solid line: Exact analytical result 
(\ref{7})-(\ref{9}) in the absence
of disorder ($\sigma =0$).
Lower and upper stars: analytical approximations 
(\ref{21}) (no disorder) and 
(\ref{24}) (with disorder), 
respectively.
For large and small $F$ (not shown) all
curves monotonically approach 
$D(F\to\infty)=D_0$ and $D(F=0)\ll D_0$.
Main conclusion: the purely periodic potential 
leads to a maximal diffusion enhancement by about a 
factor of 70 (solid line) while a tiny amount of 
disorder (insets) further boosts the
peak by a factor of 10 (dots).
}
\label{fig1}
\end{figure}

Without the disorder $W(x)$, the following rigorous results
are known (see \cite{rei01} and references therein)
\begin{eqnarray}
& & v=D_0\, [1-e^{-LF/kT}]/A\ ,\ \ D=D_0\, B/A^3 \ ,
\label{7}
\end{eqnarray}
where $D_0:=kT/\eta$ is the force-free diffusion coefficient
according to Einstein and
\begin{eqnarray}
& & A=\int_0^L \!\! \frac{dx}{L}\int_0^L \!\! dy\, e^{[V(x)-V(x-y)]/kT}
\label{8}
\\
& & B=\int_0^L \!\! \frac{dx}{L}\int_0^L \!\! dy
\int_0^L \!\! dp \int_0^L \!\! dq\,  e^{g/kT}
\nonumber
\\
& & 
g = V(x)-V(x-y)-V(x-p)+V(x+q) \ ,
\label{9}
\end{eqnarray}
see also \cite{par97,cos99,diff1,der83} for related findings.
In particular, for $F=0$ one recovers the result \cite{lif62}
\begin{equation}
D(F=0)=\frac{D_0}{C_+C_-}\ ,\ 
C_\pm=\int_0^L \frac{dx}{L}e^{\pm V(x)/kT} \ .
\label{10}
\end{equation}

Next we include the disorder $W(x)$ in (\ref{2}).
In a first step we assume that $W(x+NL)=W(x)$ for
some integer $N$. Hence $U(x)$ in (\ref{2}) is a tilted
periodic potential with period $NL$.
Accordingly, (\ref{7})-(\ref{10}) remain valid
after replacing $L$ by $NL$ and $V(x)$ by $U(x)$.
Exploiting (\ref{2}), one can rewrite (\ref{8}) 
after some manipulations as
\begin{eqnarray}
& & A=\int_0^L \!\! \frac{dx}{L}\int_0^L \!\! dy\, e^{[V(x)-V(x-y)]/kT}
\tilde A(x,y)
\label{11}
\\
& & \tilde A(x,y):=\sum_{n=0}^{N-1}\langle
e^{[-nLF+W(x)-W(x-y-nL)]/kT}\rangle
\label{12}
\end{eqnarray}
where $\langle f(x) \rangle:=N^{-1}\sum_{\nu=0}^{N-1}f(x+\nu L)$.
Next, we let $N\to\infty$ and adopt the usual tacit
assumption \cite{der83,sch95}
that this limit commutes with the limit 
$t\to\infty$ in (\ref{3}).
Exploiting that the Gaussian process $W(x)$ is ergodic
and satisfies $\langle e^W\rangle=e^{\langle W^2 \rangle/2}$
yields
\begin{eqnarray}
& & \tilde A(x,y) =\sum_{n=0}^{\infty}
e^{[-nLF + \tilde c(y+nL)]/kT}
\label{13}
\\
& & 
\tilde c(x):= [\sigma^2 -c(x)]/kT \ .
\label{14}
\end{eqnarray}
From the monotonicity of $c(x)$ (see above (\ref{5}))
one can infer upper and lower bounds
for (\ref{13}) and thus for (\ref{11}).
Finally, one recovers the same formula 
for $v$ as in (\ref{7}) but now with the 
relevant  $A$ given for any $F\geq 0$ by
\begin{eqnarray}
& &\!\!\! \!\!\!\!\!\!\!\!\! 
A=(1+a) \int_0^L \!\! \frac{dx}{L}\int_0^L \!\!
dy\, e^{[V(x)-V(x-y)+\tilde c(y)]/kT}
\label{15}
\\
& &\!\!\!\!\!\!\!\!\!\!\!\! 
a=e^{-LF/kT}\, [e^{\vartheta(\sigma/kT)^2} -1]
\label{16}
\end{eqnarray}
with some (unknown) $\vartheta\in [0,1]$.
Note that $a$ and $\tilde c(y)$ are non-negative and 
that both quantities vanish (for all $y$) if and only if
there is no disorder.
In the latter case, (\ref{15}) reproduces (\ref{8}), 
otherwise {\em the disorder always reduces the velocity $v$}.
This conclusion in fact applies to much more general types of
disorder $W(x)$, as can be inferred by applying Jensen's
inequality to (\ref{12}).
Finally, it follows from (\ref{15}) that 
$v\to F/\eta$ for $F\to\infty$,
independently of $V(x)$ and $W(x)$.
For related findings see also \cite{sch95}.

An analogous calculation yields the same formula for 
the diffusion as in (\ref{7}) with $A$ given by 
(\ref{15}) and $B$ by
\begin{eqnarray}
& &\!\!\! \!\!\!\!\!\!\!\!\! \!\!\! \!\!\! 
B=(1+b) \int_0^L \!\! \frac{dx}{L}\int_0^L \!\! dy 
\int_0^L \!\! dp \int_0^L \!\!  dq\, e^{[g+h]/kT}
\label{17}
\end{eqnarray}
for any $F\geq 0$, where $g$ is defined in (\ref{9}) and
\begin{eqnarray}
& &\!\!\! \!\!\!\!\!\!\!\!\! \!\!\! \!\!\! 
h = \tilde c(y) + \tilde c(p) - \tilde c(q) -
\tilde c(y\!-\!p) + \tilde c(y\!+\!q) + \tilde c(p\!+\!q)
\label{18}
\\
& &\!\!\! \!\!\!\!\!\!\!\!\! \!\!\! \!\!\! 
b = [1-(1-e^{-LF/kT})^3]\,  [e^{\kappa(\sigma/kT)^2}-1]
\label{19}
\end{eqnarray}
with some (unknown) $\kappa\in [-2,5]$.
One readily sees that $1+b>0$ and
that (\ref{9}) is recovered in the absence of $W(x)$.
However, whether the disorder enhances or reduces the 
diffusion is not immediately obvious, with the following
exceptions: From (\ref{7}), (\ref{15}), (\ref{17})
one finds that 
\begin{equation}
D(F\to\infty) = D_0
\label{19a}
\end{equation}
and from (\ref{10}) - analogous to the derivation of (\ref{15}) - that
\begin{equation}
D(F=0)=(D_0/C_+C_-)\, e^{-(\sigma/kT)^2} \ .
\label{20}
\end{equation}
With Cauchy-Schwartz's inequality it follows \cite{zwa88}
that $D (F=0) \leq D_0$ 
and with (\ref{10}), (\ref{20}) that
{\em the disorder always reduces the diffusion for $F=0$}.

As exemplified in Fig. 1 and discussed in detail 
in \cite{rei01}, without disorder, the 
diffusion $D$, considered as a function of the 
tilt $F$, develops a pronounced peak near
$F_c := \max_x V_0'(x)$. 
The critically tilted potential 
$V(x)=V_0(x)-xF_c$ thus exhibits 
a strictly negative slope ($V'<0$) apart from
plateaux ($V'=V''=V''_0=0$, $V'''=V_0'''<0$)
at $x_c+nL$ for a generically unique $x_c\in[0,L)$ 
and arbitrary integers $n$.
In other words, in the noiseless dynamics 
(\ref{1}), $F_c$ marks the transition from
locked to running deterministic solutions.
For finite but weak noise, the peak of 
$D(F)$ about $F_c$ satisfies \cite{rei01}
\begin{eqnarray}
D(F_c,W=0)\simeq 0.021\, D_0\, L^2\, |V_0'''(x_c)/kT|^{2/3} 
\label{21}
\end{eqnarray}
With our findings $D(F=0)\leq D_0$ and 
$D(F\to\infty)=D_0$ we thus can conclude that
the peak height of $D(F)/D_0$ scales like $T^{-2/3}$, 
and similarly for its width \cite{rei01}.

Next we consider the influence of the disorder $W(x)$
in the above most interesting regime of small 
$F-F_c$ and $kT$.
As intuitively expected and confirmed
by closer inspection, in this regime the integrals
in (\ref{15}) and (\ref{17}) are dominated by 
small values of $x-x_c$ $y$, $p$, $q$, thus 
admitting the following approximate expansions
\begin{eqnarray}
\!\!\!\!\!\!\!\!\! V(x_c+\delta) & \simeq & V (x_c)-(F\! - \!  F_c)\, \delta + V_0'''(x_c)\, \delta^3/6 
\label{22}
\\
\tilde c(x) & \simeq & \tilde c''(0)\, x^2/2 \ .
\label{23}
\end{eqnarray}
Further, we henceforth neglect $a$ and
$b$ in (\ref{15}) and (\ref{17}) \cite{f2}.
The main virtues of these approximations are:
(i) They still reproduce the correct limiting 
behavior for $F\to\infty$ and also qualitatively 
capture the small $F$ behavior, namely extremely 
small values of drift and diffusion.
In other words, they are expected to reasonably 
work for all $F\geq 0$, as confirmed by Fig. 1.
(ii) The main effects of $V_0(x)$ and $W(x)$ are 
already captured by $F_c$, $V_0'''(x_c)$, 
and $\langle W'(x)^2\rangle=-c''(0)$.
(iii) For $F=F_c$ and sufficiently small $kT$,
the integrands in (\ref{15}) and (\ref{17})
exhibit very pronounced maxima and thus can 
be evaluated by means of saddle point 
approximations, yielding the ``universal scaling law''
\begin{eqnarray}
D(F_c) & \simeq & D(F_c,W=0)\, \left(1+1.9\ Q\, e^{416 Q^3/3}\right)
\label{24}
\\
Q & := & \langle W'(x)^2\rangle/|V_0'''(x_c)|^{2/3} (kT)^{4/3} \ .
\label{25}
\end{eqnarray}

{\em The general analytical results (\ref{19a}), (\ref{20}), (\ref{24}) 
represent the main findings of our present paper:}
Already a small amount of disorder 
typically leads to a much more pronounced peak of 
$D(F)$ than without disorder. 
An illustration is provided by Fig. 1.

\begin{figure}
\epsfxsize 1.0 \columnwidth
\epsfbox{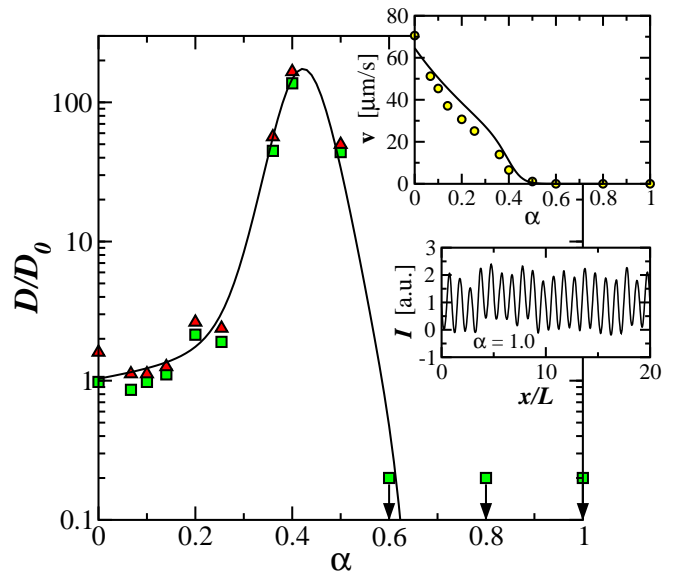}
\caption{
Symbols: Experimentally measured diffusion $D$ and 
velocity $v$ (upper inset), adopted from Fig. 3 of Ref. \cite{lee06}.
Solid lines: Theoretical fit.
Lower inset: Approximate intensity profile
in arbitrary units for $\alpha=1.0$ and 
$x \in [0,20 L]$. 
}
\label{fig2}
\end{figure}

As an application of our general theory, 
we finally address the experiment from
Ref. \cite{lee06}:
A colloidal sphere with diameter $1.48\,\mu$m and 
$D_0=kT_{room}/\eta \simeq 0.19\,\mu$m$^2/$s
moves along a ring of light.
The particle feels $N=80$ potential minima with period 
$L\simeq 0.33\,\mu$m due to spatial variations of
the light intensity and a torque due to orbital 
angular momentum transfer by the photons
from two superimposed optical
vortices \cite{lee06,cur03}, 
whose relative strengths are controlled by an 
experimental parameter $\alpha \in [0,1]$.
The resulting equation of motion takes the form 
(\ref{1}) \cite{cur03} with a potential (\ref{2}) 
corresponding to the total circumferential force 
from Eq. (4) in \cite{lee06}, namely
\begin{equation}
-U'(x) =  F_0 [\Phi (\alpha) +  \Psi (\alpha)\,\cos(2\pi x/L)] - W'(x) 
\label{26}
\end{equation}
with $\Phi (\alpha):=(1-\alpha)/(1+\alpha)$ and 
$\Psi  (\alpha) := 2[\alpha (\epsilon^2 \Phi ^2+\zeta^2)]^{1/2}/(1+\alpha)$.
Further, $F_0$, $\epsilon$, and $\zeta$
are fit parameters, accounting for the laser intensity 
and the particle's shape, size, and composition 
(light scattering and absorption properties).
The torque being proportional to the light intensity
gives rise to the first term
on the right hand side of (\ref{26}) (fit parameter $F_0$) 
and also to part of the second term \cite{cur03}
(fit parameter $\epsilon$).
Additionally, the second term accounts 
for the polarizable particle's coupling to the
gradient of the light intensity
(fit parameter $\zeta$).
Finally, $W(x)$ in (\ref{26}) accounts for 
``random'' imperfections of the experimental optics \cite{lee06}.
Its variance is fixed by observing that
$-W'(x)/F_0$ corresponds to the function
$\eta (\theta)$ in \cite{lee06} and that
$\langle |\eta(\theta)|^2\rangle \approx 0.01$ 
according to \cite{lee06}.
Further statistical properties of $W(x)$
cannot be quantitatively related to those of 
the bare intensity reported in 
\cite{lee06} since both the intensity and the 
intensity gradient contribute -- after suitably 
averaging over the particle volume -- to $W(x)$.
For this reason, we model $W(x)$ as Gaussian 
process (\ref{6}) with periodicity $W(x+NL)=W(x)$ and 
$\langle W'(x)^2\rangle = \sigma^2/\lambda^2 = 0.01\,F_0^2$.
Regarding the correlation length $\lambda$,
we found that its exact quantitative value
hardly matters and we have chosen 
$\lambda=2L$, in accordance with 
both Fig. 1d from \cite{lee06} 
and the given particle size.
Fig. 2 depicts our fit to the experimental 
results with parameter values 
$F_0=1.37$ pN, $\epsilon=0.38$, and $\zeta=0.25$.
\cite{par}.
In view of the periodicity $W(x+NL)=W(x)$
we used the exact analytics (\ref{7}) 
with $A$ from (\ref{11}) and similarly 
for $B$.
The remaining dependence of the results on
the realization of $W(x)$ is still notable
for $N=80$.
In the absence of more than one experimental 
realization, we have selected also in the 
theory a well fitting,  but still 
representative single realization 
$W(x)$.
The minor differences between
theory and experiment can be naturally
attributed to the fact that this 
$W(x)$ is still not exactly the one
realized in the experiment and to 
oversimplifications of the theoretical 
model (\ref{1}) {\em per se}.
Once $W(x)$ and all fit parameters
are fixed in (\ref{26}), 
it is possible to approximately
estimate the underlying bare 
intensity $I(x)$.
The resulting $I(x)$ in Fig. 2
indeed agrees quite well with
Fig. 1d from \cite{lee06}.
All in all, our theory thus agrees in 
every respect very well with the 
experiment from \cite{lee06}.
An analogous comparison with the experiments
from \cite{bli07,evs08} is prohibited by their 
small $N$-values. 

In conclusion, our main finding consists 
in the general analytical results 
(\ref{19a}), (\ref{20}), (\ref{24}),
implying that even a tiny amount of disorder 
superimposed to the (dominating) periodic 
potential may further boost the previously 
known sharp diffusion peak near the critical 
tilt by orders of magnitude, see Fig. 1.
A further main point of our analytical 
findings is the universality of this very 
selective and very strong diffusion 
enhancement close to the critical tilt. 
Considering that different species of particles
typically couple differently to the periodic and
random potential and/or to the bias force 
we expect different values of the critical
tilt for each species.
This opens the possibility of sorting 
particles by way of selectively enhancing 
the diffusion for certain species within 
a mixture.
The experimentally often unavoidable weak 
disorder quite unexpectedly improves rather 
than deteriorates the effectivity of the
selection mechanism.
Experiments along these lines for a mixture of
different DNA-fragments in a periodically structured 
microfluidic device analogous to those 
from \cite{dna} are presently in preparation
in the labs of D. Anselmetti at the University 
of Bielefeld.

\begin{center}
\vspace{-1mm}
---------------------------
\vspace{-5mm}
\end{center}
This work was supported by the Deutsche 
Forschungsgemeinschaft under SFB 613.


\begin{thebibliography}{10}
\bibitem{f2}
A sufficient condition to do so is
$F_cL/kT-5(\sigma/kT)^2\gg 1$, 
see (\ref{16}), (\ref{19}).
Closer inspection yields the weaker condition
$F_cL/kT-3(\tilde \sigma/kT)^2\gg 1$ where 
$\tilde\sigma^2:=\max_{x,n}|c(x+nL)-c(x)|/n$.
\bibitem{par}
$F_0$ mainly affects $v(\alpha)$ for small $\alpha$,
$\epsilon$ the shape, and $\zeta$ the position of the 
$D(\alpha)$-peak. 
Changing them by 10-20\% still yielded 
satisfactory results, and also the detailed
properties of $W(x)$ turned out 
to be quite uncritical.
\end{thebibliography}
\end{document}